\documentclass[11pt]{elsarticle}
\usepackage{mhchem}
\usepackage[utf8]{inputenc}
\usepackage[T1]{fontenc}
\usepackage{xspace}
\usepackage{fullpage}
\usepackage{graphicx}
\usepackage{lmodern}
\usepackage{algorithm2e}
\usepackage{framed}
\usepackage{multirow}
\usepackage{tikz}
\usepackage{pgfplots}

\usepackage{amsmath,amsfonts,amssymb}

\DeclareMathOperator*{\argmin}{arg\,min}

\newcommand{\lela}{\left \langle}
  \newcommand{\rira}{\right \rangle}
\newcommand{\norm}[1]{\left\lVert#1\right\rVert}

\newcommand{\axisopt}{height = 10cm, width = 14cm}

\usepackage{algorithmic}
\usepackage{hyperref}

\newcommand{\npw}{\ensuremath{N_{\text{pw}}}\xspace}
\newcommand{\nline}{\ensuremath{n_{\text{inner}}}\xspace}
\newcommand{\nprojs}{\ensuremath{N_{\text{projs}}}\xspace}
\newcommand{\nband}{\ensuremath{N_{\text{bands}}}\xspace}
\newcommand{\nblocks}{\ensuremath{N_{\text{blocks}}}\xspace}
\newcommand{\nppw}{\ensuremath{p_{\text{pw}}}\xspace}
\newcommand{\npband}{\ensuremath{p_{\text{bands}}}\xspace}
\newcommand{\natom}{\ensuremath{N_{\text{atoms}}}\xspace}
\newcommand{\nlmn}{\ensuremath{n_{lmn}}\xspace}

\definecolor{line01}{RGB}{242, 201,  49}
\definecolor{line02}{RGB}{ 33, 110, 180}
\definecolor{line03}{RGB}{154, 153,  64}
\definecolor{line04}{RGB}{187,  77, 152}
\definecolor{line05}{RGB}{222, 139,  83}
\definecolor{line06}{RGB}{121, 187, 146}
\definecolor{line07}{RGB}{223, 154, 177}
\definecolor{line08}{RGB}{197, 163, 202}
\definecolor{line09}{RGB}{205, 200,  63}
\definecolor{line10}{RGB}{223, 176,  57}
\definecolor{line11}{RGB}{142, 101,  56}
\definecolor{line12}{RGB}{ 50, 142,  91}
\definecolor{line13}{RGB}{137, 199, 214}
\definecolor{line14}{RGB}{103,  50, 142}

\definecolor{linea}{RGB}{211,  94,  60}
\definecolor{lineb}{RGB}{ 78, 144, 204}

\usepackage{xcolor}
\colorlet{chebcolor}{linea}
\colorlet{lobpcgcolor}{lineb}
\colorlet{cgcolor}{line14}
\newcommand{\chebmarker}{square}
\newcommand{\lobpcgmarker}{triangle}

\begin{document}
\begin{frontmatter}

  \begin{abstract}
    We consider the problem of parallelizing electronic structure
    computations in plane-wave Density Functional Theory. Because of
    the limited scalability of Fourier transforms, parallelism has to
    be found at the eigensolver level. We show how a recently proposed
    algorithm based on Chebyshev polynomials can scale into the tens
    of thousands of processors, outperforming block conjugate gradient
    algorithms for large computations.
  \end{abstract}
  \title{Parallel eigensolvers in plane-wave Density Functional Theory}
  \author{Antoine Levitt\corref{cor1}}
  \ead{antoine.levitt@gmail.com}

  \author{Marc Torrent}
  \ead{marc.torrent@cea.fr}

  \cortext[cor1]{Corresponding author}
  \address{CEA, DAM, DIF, F-91297, Arpajon, France}
\end{frontmatter}
\tableofcontents

\section{Introduction}

Kohn-Sham Density Functional Theory is an efficient way to solve the
Schrödinger equation for quantum systems
\cite{hohenberg1964inhomogeneous, kohn1965self}. By modelling the
correlation between $N$ electrons via exchange-correlation
functionals, it leads to the Kohn-Sham system, mathematically
formulated as a nonlinear eigenvalue problem. This problem can be
discretized and solved numerically, and the result of this
computation allows the determination of physical properties of
interest via higher-level processing such as geometry optimization,
molecular dynamics or response-function computation. Density
Functional Theory (DFT) codes can be classified according to the
discretization scheme used to represent wavefunctions (plane waves,
localized basis functions, finite differences ...) and the treatment
of core electrons (all-electron computations, pseudopotentials
...). We focus on the ABINIT software \cite{gonze2009abinit}, which
uses a plane-wave basis and either norm-conserving pseudopotentials or
the Projector Augmented-Wave (PAW) approach
\cite{blochl1994projector,torrent2008implementation}.

The bottleneck of most simulations is the computation of the
electronic ground state. This is done by a self-consistent cycle whose
inner step is the solution of a linear eigenvalue problem. This step
has to be implemented efficiently, taking into account the
specificities of the problem at hand, which rules out the use of
generic black-box solvers. Furthermore, the growing need for
parallelization constrains the choice of the eigensolver. Indeed, one
specificity of plane-wave DFT as opposed to real-space codes is that
Fourier transforms do not scale beyond about 100 processors: effective
parallelization requires eigensolvers that are able to compute several
Hamiltonian applications in parallel.

The historic eigensolver used in plane-wave DFT, the conjugate
gradient scheme of refs. \cite{payne1992iterative,
  kresse1996efficient}, is inherently sequential, although there are
attempts at parallelization by omitting orthogonalizations
\cite{iwata2010massively}. Several methods work on blocks of
eigenvectors and are more suited for parallelization, such as the
residual vector minimization -- direct inversion in the iterative
subspace (RMM-DIIS) scheme \cite{kresse1996efficient}, and block
Davidson algorithms \cite{davidson1975iterative}, including the
locally optimal block preconditioned conjugate gradient (LOBPCG)
algorithm \cite{knyazev2001toward}, implemented in ABINIT
\cite{bottin2008large}.

Parallel implementations of plane-wave DFT codes include Quantum
Espresso \cite{giannozzi2009quantum}, VASP \cite{kresse1996efficient},
QBOX \cite{gygi2006large} or CASTEP \cite{milman2010electron}.  The
scalability of these codes is mainly limited by orthogonalizations and
the Rayleigh-Ritz step, a dense matrix diagonalization, which is hard
to parallelize efficiently, even using state-of-the-art libraries such
as ELPA \cite{auckenthaler2011parallel} or Elemental
\cite{poulson2013elemental}. The Rayleigh-Ritz step usually becomes
the bottleneck when using more than a thousand processors.

There are two main ways to decrease the cost of this step. One is to
use it as rarely as possible. This usually means applying the
Hamiltonian more than one time to each vector before applying the
Rayleigh-Ritz procedure, in order to speed up convergence. The other
is getting rid of it entirely. This requires the independent
computation of parts of the spectrum, as in the methods of spectrum
slicing \cite{slicing} or of contour integrals \cite{polizzi,
  sakurai2003projection}. These approaches effectively solve an
interior eigenvalue problem, which is considerably harder than the
original exterior one. The result is that a large number of
Hamiltonian applications is needed, to obtain a high-degree polynomial
or to solve linear systems.

While these spectrum decomposition techniques will surely become the
dominant methods for exascale computing, we address the current
generation of supercomputers, on which the decrease in the costs of
the Rayleigh-Ritz step is not worth the great increase in the number
of Hamiltonian applications. We therefore focus in this paper on the
method of Chebyshev filtering, which aims to limit the number of
Rayleigh-Ritz steps by applying polynomials of the Hamiltonian to each
vector. It can be seen as an accelerated subspace iteration, and dates
back to the RITZIT code in 1970 \cite{rutishauser1970simultaneous}. It has
been proposed for use in DFT in refs. \cite{seq_chebfi,
  parallel_chebfi}, and has recently been adopted by several groups
\cite{berljafa2014optimized, banerjee2014spectral}.

The contribution of this paper is twofold. First, we show how to adapt
the Chebyshev filtering algorithm of ref. \cite{seq_chebfi} in the
context of generalized eigenproblems, here due to the PAW
formalism. By exploiting the particular nature of the PAW overlap
matrix (a low-rank perturbation of the identity), we are able to
invert it efficiently. Second, we compare the Chebyshev filtering
algorithm with CG and LOBPCG, both in terms of convergence and
scalability.

\section{The eigenvalue problem}
\subsection{The operators}
First, we define some relevant variables. For a system of $\natom$
atoms in a box, we solve the Kohn-Sham equations in a plane-wave basis. This
basis is defined by the set of all plane waves whose kinetic energy is less
than a threshold $E_{\text{cut}}$. This yields a sphere of $\npw$
plane waves, upon which the wavefunctions are discretized.

We consider a system where \nband bands are sought. For a simple
ground state computation, \nband represents the number of states
occupied by valence electrons of the $\natom$ atoms. For more
sophisticated analysis such as Many-Body Perturbation Theory (MBPT),
the computation of empty states is necessary, and \nband can be
significantly higher. It is also convenient to speed up convergence of
ground state computations to use more bands than strictly necessary.

To account for the core electrons, we use pseudopotentials. ABINIT
implements both norm-conserving pseudopotentials and the Projector
Augmented-Wave (PAW) method. For the purposes of this paper, the main
difference is the presence of an overlap matrix in the PAW case,
leading to a generalized eigenvalue problem. We will assume in the
rest of this paper that we use the PAW method: norm-conserving
pseudopotentials follow as a special case.

For simplicity of notation, we consider in this paper the case where
periodicity is not taken explicitely into account, and the
wavefunctions will be assumed to be real. The following discussion
extends to the periodic case by sampling of the Brillouin zone,
provided that we consider complex eigenproblems, with the necessary
adjustments.

The Kohn-Sham equations for the electronic wavefunctions $\psi_{n}$ are
\begin{align}
  \label{KS}
  H \psi_{n} = \lambda_{n} S \psi_{n},
\end{align}
where $H$ is the Hamiltonian, and $S$ the overlap matrix arising from
the PAW method ($S = I$ with norm-conserving pseudopotentials). $H$
and $S$ are \npw $\times$ \npw Hermitian matrices (although they are
never formed explicitely), and $\Psi$ is a \npw $\times$ \nband matrix
of wavefunctions. The Hamiltonian operator depends self-consistently
on the wavefunctions $\Psi$. It can be written in the form
\begin{align}
  H = K + V_{\text{loc}} + V_{\text{nonloc}}.
\end{align}

The kinetic energy operator $K$ is, in our plane wave basis, a simple
diagonal matrix. The local operator $V_{\text{loc}} = V_{\text{ext}} +
V_{\text{H}} + V_{\text{XC}}$ is a multiplication in real space by a
potential determined from atomic data and the wavefunctions $\Psi$. It can therefore
be computed efficiently using a pair of inverse and direct FFTs. The
nonlocal operator $V_{\text{nonloc}}$ and the overlap matrix $S$
depend on the atomic data used. For both PAW method and
norm-conserving pseudopotentials, we introduce a set of
$\nlmn$ projectors per atom, where $\nlmn$ is the
number of projectors used to model the core electrons of each atom,
and usually varies between 1 and 40 according to the atom and
pseudopotential type. Therefore, for a homogeneous system of $\natom$
atoms we use a total of $\nprojs = \nlmn \natom$
projectors. We have $\nprojs \ll \npw$, but \nprojs is comparable to
\nband.

We gather formally these projectors in a \npw$ \times$ \nprojs matrix
$P$. The non-local operator $V_{\text{nonloc}}$ is computed as
\begin{align}
  V_{\text{nonloc}} = P D_{V} P^{T}.
\end{align}

Similarly, the overlap matrix in the PAW formalism is
\begin{align}
  S = I + P D_{S} P^{T}.
\end{align}

The matrices $D_{S}$ and $D_{V}$ do not couple the different atoms in
the system: they are block-diagonal. They can be precomputed from
atomic data. The matrix $D_{V}$ additionally depends self-consistently
on the wavefunctions $\Psi$.

Therefore, for a single band $\psi$, the process of computing
$H \psi$ and $S \psi$ can be decomposed as follows

\begin{framed}

  \begin{algorithm}[H]
    \caption{Computation of $H \psi, S \psi$}
    \begin{algorithmic}
      \STATE \textbf{Input:} a wavefunction $\psi$
      \STATE \textbf{Output:} $H\psi$, $S \psi$
      \STATE $\circ$ Compute $K \psi$ by a simple scaling
      \STATE $\circ$ Apply an inverse FFT to $\psi$ to compute its real-space
      representation, multiply by $V_{\text{loc}}$,

      and apply a FFT
      to get $V_{\text{loc}} \psi$
      \STATE $\circ$ Compute the \nprojs projections $p_{\psi} = P^{T} \psi$
      \STATE Apply the block-diagonal matrices $D_{V}$ and $D_{S}$
      to $p_{\psi}$
      \STATE Compute the contributions $P D_{V} p_{\psi}$ and $P
      D_{S} p_{\psi}$ to the nonlocal and overlap operator
      \STATE $\circ$ Assemble $H \psi = K \psi + V_{\text{loc}} \psi + P
      D_{V} p_{\psi}$
      \STATE $\circ$ Assemble $S \psi = \psi + P D_{S} p_{\psi}$
    \end{algorithmic}
    \label{alg:h}
  \end{algorithm}
\end{framed}
The total cost of this operation is $O(\npw \log \npw + \npw
\nprojs)$. As $\npw$ and $\nprojs$ both scale with the number of atoms
$\natom$, the cost of computing the non-local operator dominates for
large systems. However, \npw is usually much greater than \nprojs, and
the prefactor involved in computing FFTs is much greater than the one
involved in computing the simple matrix products $P^{T} \psi$ and $P
p_{\psi}$ (which can be efficiently implemented as a level-3 BLAS
operation). The FFT and non-local operator costs are usually of the
same order of magnitude for systems up to about 50 atoms.
\subsection{Solving the eigenvalue problem: conjugate gradient}
The historical algorithm used to compute the \nband first eigenvectors
of \eqref{KS} in the framework of plane-wave DFT is the conjugate
gradient algorithm, described in \cite{payne1992iterative,
  kresse1996efficient}. It is mathematically based on the following
variationnal characterization of the $n$-th eigenvector of the
eigenproblem $H \psi = \lambda S \psi$:
\begin{align*}
  \psi_{n} &= \argmin_{\lela \psi_{i}, S \psi\rira = \delta_{i,n},
    \; i = 1,\dots,\npw}
  \lela \psi, H \psi\rira.
\end{align*}

The conjugate gradient method of ref. \cite{payne1992iterative,
  kresse1996efficient} consists of minimizing this functional by a
projected conjugate gradient method. Note that, because of the
constraints, this is a nonlinear conjugate gradient problem, to which
classical (linear) results can not be applied. A number of desirable
characteristics have made it the algorithm of reference.

First, this algorithm only needs the application of the operators $H$
and $S$ to wavefunctions, and can therefore be decoupled from their
underlying structure. This is particularly suited to plane-wave
Density Functional Theory, where the application of $H$ can be
efficiently computed with FFTs.

Second, the operator $H = K + V_{\text{loc}} + V_{\text{nonloc}}$ is
closely approximated by $K$ in the high-frequency regime. A good
diagonal preconditionner can therefore be built by damping the high
frequencies as $K^{-1}$ above a certain threshold. The implementation
in \cite{payne1992iterative, kresse1996efficient}, still used in most
plane-wave codes, uses a smooth rational function with a variable
threshold (typically taken to be the kinetic energy of the band under
consideration). The use of a preconditionner greatly accelerates the
convergence with only a negligible additional cost of $O(\npw)$.

Third, the algorithm can naturally reuse approximate eigenvectors, in
contrast to algorithms based on a growing basis such as the Lanczos
iteration. This is extremely attractive in DFT computations, where
very good approximations can be obtained from the previous self-consistent cycle.

Typically, this algorithm is implemented in the following way: for
each band in ascending order, do a fixed number \nline of iterations
of the conjugate gradient algorithm, orthogonalizing at each step with
respect to the other bands. Once every band is updated, use a
Rayleigh-Ritz step (also called \textit{subspace rotation} or
\textit{subspace diagonalization}), update the density (usually, Pulay
mixing with preconditionning is used), and iterate until
convergence. Therefore, one has a system of inner-outer iterations
controlled by the variable \nline. To our knowledge, little is known
about the correct way to choose this parameter, especially if it is
allowed to vary between bands.

\subsection{Block algorithms: LOBPCG}
A number of alternative approaches have been developed over the
years. We focus in this section on the Locally-Optimal Block
Preconditioned Conjugate Gradient (LOBPCG) algorithm
\cite{knyazev2001toward}, which was developed as a way to improve the
convergence of the conjugate gradient method. In its single-block
version, it consists of a Rayleigh-Ritz method in the
$3\nband$-dimensional subspace spanned by the current trial
wavefunctions, the wavefunctions computed at the previous iteration,
and the (preconditionned) residuals. For a single band, this would be
equivalent to the conjugate gradient algorithm. For multiple bands, by
computing the eigenvectors as the solution to an eigenvalue problem in
a well-chosen subspace, this method achieves higher convergence rates
\cite{knyazev2001toward}.

The price to pay for this faster convergence is the solution of a
dense eigenvalue problem of size $3\nband$ for the Rayleigh-Ritz
method. While this cost is negligible for small systems, where the
cost of applying the Hamiltonian dominates the computation, its
$O(\nband^{3})$ cost becomes problematic for larger systems,
especially since it has poor parallel scaling. For this reason, a
``multiblock'' scheme has been implemented in ABINIT
\cite{bottin2008large}. In this scheme, the $\nband$ bands are split
in $\nblocks$ blocks. The LOBPCG algorithm is applied in each block,
which is additionally kept orthogonal to the blocks of lower
energy. In this way, the Rayleigh-Ritz cost is cut by a factor
$\nblocks^{3}$. The LOBPCG algorithm as described suffers from
ill-conditioning of the matrices involved in the Rayleigh-Ritz step,
and practical implementations have to be modified to use a more
suitable basis (see \cite{knyazev2001toward, hetmaniuk2006basis}), but
the resulting method has proven robust and improves the convergence of
the conjugate gradient algorithm.

The main advantage, and motivation of its adoption in ABINIT, is
however not its improved convergence, but the ability to build the
residuals for all the bands of a block in parallel, as will be
discussed in the next section.
\section{Parallelism}
We consider the parallelization of ground state computations in
plane-wave DFT, and its implementation in ABINIT.

\subsection{Parallelism in the Hamiltonian application}
The most straightforward way of parallelizing problem \eqref{KS} is to
use multiple processors to compute the Hamiltonian application $H
\psi$ needed in the conjugate gradient algorithm. In this approach,
the vector $\psi$ of size \npw is distributed onto \nppw processors,
and the Hamiltonian is computed in parallel. Although the non-local
part can be computed very efficiently in this approach, the parallel
computation of FFTs is a challenge. 3D FFT can be parallelized by
computing multiple 2D FFTs in parallel, but this approach is
intrinsically unable to exploit more than $\npw^{1/3}$
processors. Even for large systems, with $\npw$ of about 1 million,
this only amounts to using 100 processors, which is clearly
insufficient to use today's supercomputers.

Therefore, in contrast with codes that work in a real-space localized
basis, our delocalized basis is an obstacle to parallelism, limiting
the scaling of the Hamiltonian application. To be used efficiently on
supercomputers, parallelism must be found elsewhere.
\subsection{Eigenvector-level parallelism in block algorithms}
Another way of solving \eqref{KS} in parallel is to use a block
algorithm such as LOBPCG, in which the Hamiltonian application on the
different vectors inside a block is done in parallel. This is the
approach taken by ABINIT.

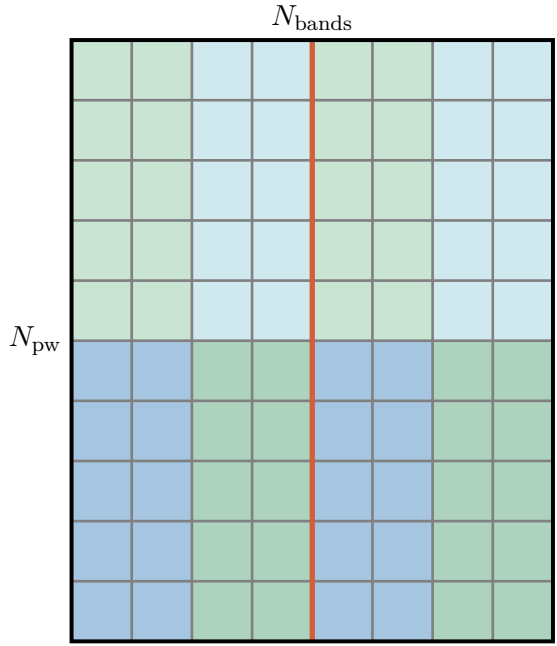
\begin{figure}[h!]
  \centering
  \begin{tikzpicture}[scale=.8]

    \pgfmathsetmacro\lx{4}
    \pgfmathsetmacro\ly{10}

    \definecolor{m1}{named}{line02}
    \definecolor{m2}{named}{line06}
    \definecolor{m3}{named}{line12}
    \definecolor{m4}{named}{line13}
    \definecolor{lines}{named}{gray}

    \pgfmathsetmacro\shift{40}

    \foreach \x/\y/\clr in {0/0/m1,0/1/m2,1/0/m3/,1/1/m4/} {
      \fill[color=\clr!\shift] (0.5*\lx*\x,0.5*\ly*\y) rectangle ({0.5*\lx*(\x+1)},{0.5*\ly*(\y+1)});
    }
    \foreach \x/\y/\clr in {2/0/m1,2/1/m2,3/0/m3/,3/1/m4/} {
      \fill[color=\clr!\shift] (0.5*\lx*\x,0.5*\ly*\y) rectangle ({0.5*\lx*(\x+1)},{0.5*\ly*(\y+1)});
    }

    \foreach \x in {2,3,...,8} {\draw[color=lines,line width=1pt] (\x-1,0) -- (\x-1,\ly);}
    \foreach \y in {2,3,...,\ly} {\draw[color=lines,line width=1pt] (0,\y-1) -- (2*\lx,\y-1);}
    \draw[color=linea,line width=2pt] (\lx,0) -- (\lx,\ly);
    \draw[color=black,line width=1.5pt] (0,0) rectangle (2*\lx,\ly);
    \node[left] at (0,\ly/2) {\npw};
    \node[above] at (\lx,\ly) {\nband};
  \end{tikzpicture}
  \caption{Distribution of the wavefunctions $\Psi$, with $\nband = 8$,
    $\npw = 10$. The data is distributed on a 2D processor grid of $\nppw = 2$,
    $\npband = 2$. Each processor is in charge of $\nblocks = 2$ blocks
    of size $\npw / \nppw = 5$ by $\nband / \npband / \nblocks = 2$.}
  \label{fig:distribution}
\end{figure}

In this approach, the wavefunction matrix $\Psi$ is distributed along
a 2D grid of $\nppw \times \npband$ processors (see Figure
\ref{fig:distribution}). The Hamiltonian can be applied with only
column-wise communications between $\nppw$ processors. The
orthogonalization and Rayleigh-Ritz steps are done using a
transposition to a $(\nppw \npband) \times 1$ processor
grid. Implementation details can be found in \cite{bottin2008large}.

Compared to using parallelism only in the Hamiltonian application, the
scalability of the code is greatly extended, up to hundreds and even
thousands of processors for large systems. The main obstacle to
parallelism is the poor scalability of the Rayleigh-Ritz (subspace
diagonalization) step. Even using parallel solvers such as ScaLAPACK
\cite{choi1992scalapack} or ELPA \cite{auckenthaler2011parallel}, the
diagonalization stops scaling at around 100 processors for large
systems, and becomes the bottleneck when using many
processors. Furthermore, as the blocksize has to be at least equal to
\npband, as the number of processors increase, so does the size of the
intermediate Rayleigh-Ritz procedures, quickly degrading performance.

To reduce these costs, we should ideally get rid of the global
Rayleigh-Ritz step. This requires computing parts of the spectrum in
parallel, which means solving an interior eigenvalue problem,
requiring considerably more matrix-vector operations. An intermediate
approach is to limit the number of global operations, i.e. to use
several matrix-vector application for each Rayleigh-Ritz step.

\section{Chebyshev filtering}
\subsection{Filtering algorithms}
The filtering approach to eigenvalue problems emphasizes the invariant
subspace spanned by the eigenvectors rather than the individual
eigenvectors and eigenvalues. To obtain this invariant subspace, one
uses a \textit{filter}, an approximation of the spectral projector on
the invariant subspace. Starting from an approximation to a basis of
the invariant subspace, one applies the filter to each vector. Then,
the basis is orthonormalized to prevent instability, and the process
is iterated until convergence. Once a basis of the invariant subspace is
obtained, a Rayleigh-Ritz procedure can be applied to recover the
individual eigenvectors and eigenvalues.

This procedure can be seen as an accelerated version of the classical
subspace iteration algorithm. This method is generally considered
inferior to Krylov methods such as the Lanczos algorithm, but has a
number of advantages that make it attractive in our context. First, it
is able to use naturally the information of previous self-consistent
iterations. Second, the filtering step can be done in parallel on each
vector, with interaction between vectors only occuring in the Rayleigh-Ritz
phase.

Another motivation for the use of filtering algorithms (see for
instance \cite{bekas2005computing}) is that, in many cases, one does
not need the individual eigenvectors and eigenvalues, but aggregate
quantities such as the density, that can be computed from any
orthonormal basis. Therefore, one can avoid the Rayleigh-Ritz step
altogether. We do not exploit this for two reasons. First, while this
approach does avoid the dense diagonalization in the Rayleigh-Ritz
step, it still requires an orthogonalization, which also scales
poorly. Second, the algorithm becomes less stable, and provides less
opportunities for error control (such as residuals) and
locking. Third, it is not obvious how to accomodate occupation
numbers, which, because of smearing schemes employed in computations
of metals, depend self-consistently on the eigenvalues.

Several forms of filters have been proposed in the
literature. \cite{polizzi} and \cite{sakurai2003projection} both use
rational filters originating from discretizations of contour integrals
to approximate the spectral projector. This is very efficient,
provides numerous opportunities for parallelization and has the
advantage of yielding ``flat'' filters, which have better stability
properties. It is however inefficient in our case because it requires
inversions of systems of the form $(H - z S) x = b$, where $z$ is a
complex shift. This becomes very poorly conditionned when the shift
$z$ becomes close to the real axis. Since our matrix is not sparse,
one cannot rely on factorizations to solve these systems, and our
tests have shown that solving these systems using iterative methods is
too slow to be competitive.

Restricting ourselves to only Hamiltonian applications yields
polynomial filters, that are less efficient but faster to compute than
rational filters. Since we are looking for a filter that is minimal on
the unwanted part of the spectrum, the natural idea is to use
Chebyshev polynomials, as proposed in ref. \cite{seq_chebfi}. This is
the approach we take in this paper.



\begin{figure}[h!]
  \centering
  {
    \begin{tikzpicture}
      \begin{axis}[xmin=-2,xmax=3,ymin=-0.2,ymax=1.2, legend entries =
        {Exact, Chebyshev, FEAST}, legend style = {at={(0.7,0.9)}, anchor= north
          west}, legend cell align=left, \axisopt]
        \addplot[mark=none, ultra thick] table [x index = 0, y index = 1 ] {filters_data.txt};
        \addplot[mark=none,draw=chebcolor!110,loosely dashed, line width =2] table [x index = 0, y index = 2 ] {filters_data.txt};
        \addplot[mark=none,draw=lobpcgcolor,densely dotted, line width = 2] table [x index = 0, y index = 3 ] {filters_data.txt};
      \end{axis}
    \end{tikzpicture}
  }
  \caption{Approximate filters to compute the $[-1, 0]$ part of the
    full spectrum $[-1, 2]$. The Chebyshev polynomial is of degree 4,
    FEAST corresponds to the rational approximation of ref. \cite{polizzi}
    with 8 quadrature points (with symmetry, this amounts to 4 linear solves).}
  \label{fig:filters}
\end{figure}
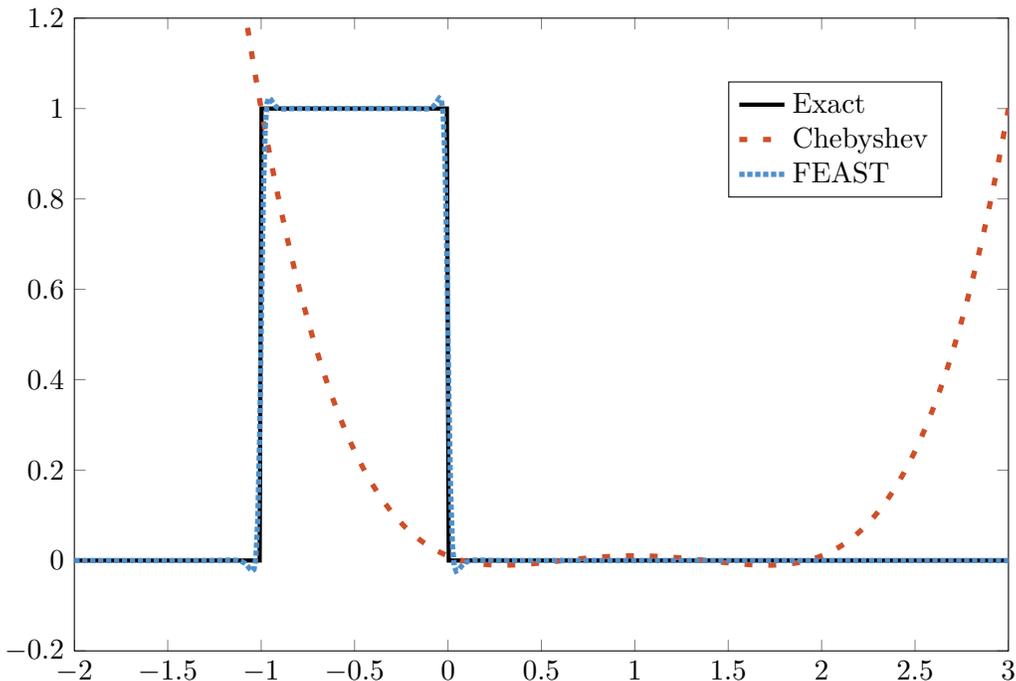

\subsection{Chebyshev filtering}
The Chebyshev polynomials have the property of being minimal in
$L^{\infty}$ norm on an interval $[a, b]$ among the polynomials of fixed degree
and scaling. They are defined recursively by
\begin{align*}
  T_{0}(x) &= 1,\\
  T_{1}(x) &= \frac {x-c}r,\\
  T_{n+1}(x) &= 2 \frac {x-c}r T_{n}(x) - T_{n-1}(x),
\end{align*}
where the filter center and radius are defined by
\begin{align*}
  c = \frac {a+b}2,\\
  r = \frac {b-a}2.
\end{align*}
This definition extends to any operator $A$ and allows us to compute $T_{n}(A)
\psi$ using $n$ applications of $A$, and with only a modest additional
memory cost.

If we denote by $\Lambda$ and $P$ the eigenvalues and eigenvectors of
the eigenproblem $H\psi = \lambda S \psi$, then we have the
decomposition $H P = S P \Lambda$, or $S^{-1} H = P \Lambda
P^{-1}$. Therefore, $T_{n}(S^{-1}H) \psi = P T_{n}(\Lambda) P^{-1}
\psi$ will have its eigencomponents filtered by the spectral filter
$T_{n}$. We then use a Rayleigh-Ritz procedure to separate the
individual eigenvectors and eigenvalues, and iterate until
convergence, as summarized in Algorithm \ref{alg:chebyshev}.

\begin{framed}
  \begin{algorithm}[H]
    \caption{Chebyshev filtering}
    \begin{algorithmic}
      \STATE \textbf{Input:} a set of $\npw \times \nband$
      wavefunctions $\Psi$
      \STATE \textbf{Output:} the updated  wavefunctions $\Psi$
      \STATE $\circ$ Compute Rayleigh quotients for every band, and set
      $\lambda_{-}$ equal to the largest one.
      \STATE $\circ$ Set $\lambda_{+}$ to be an
      upper bound of the spectrum.
      \STATE $\circ$ Compute the filter center and radius $c =
      \frac {\lambda_{+}+\lambda_{-}}2$, $r = \frac
      {\lambda_{+}-\lambda_{-}}2$

      \FOR {each band $\psi$}
      \STATE Set $\psi^{0} = \psi$, and $\psi^{1} = \frac 1 r (S^{-1} H \psi^{0} - c \psi^{0})$
      \FOR {$i = 2, \dots, \nline$}
      \STATE $\psi^{i} = \frac 2 r (S^{-1} H \psi^{i-1} - c \psi^{i-1}) - \psi^{i-2}$
      \ENDFOR
      \ENDFOR
      \STATE $\circ$ Compute the subspace matrices $H_{\psi} = \Psi^{T} H \Psi$,
      and $S_{\Psi} = \Psi^{T} S \Psi$
      \STATE $\circ$ Solve the dense generalized eigenproblem $H_{\Psi} X = S_{\Psi} X
      \Lambda$, where $\Lambda$ is a diagonal matrix

      of eigenvalues, and $X$ is the $S_{\psi}$-orthonormal set of
      eigenvectors
      \STATE $\circ$ Do the subspace rotation $\Psi
      \leftarrow \Psi X$
    \end{algorithmic}
    \label{alg:chebyshev}
  \end{algorithm}
\end{framed}

This algorithm is identical to the one found in \cite{seq_chebfi},
except that, since we are dealing with a generalized eigenproblem, we
need to apply a polynomial in $S^{-1} H$ instead of simply $H$. This
operator is not Hermitian, but has the same spectrum as the pencil
$(H,S)$, and the filtering algorithm finds the same eigenvectors and
eigenvalues with the same convergence properties as in the Hermitian
case. We will explain how to compute $S^{-1}$ efficiently in Section
\ref{sec:invovl}.

\section{Implementation}

\subsection{Inversion of the overlap matrix}
\label{sec:invovl}
We need to compute the operator $S^{-1}$, where the overlap matrix $S$
is given by
\begin{align*}
  S = I + P D_{S} P^{T}.
\end{align*}

This matrix is too large to invert directly, and is not even
sparse. However, since $\nprojs \ll \npw$, it is a low-rank
perturbation of the identity. Therefore, we can apply the Woodbury
formula \cite{woodbury1950inverting} and write its inverse as
\begin{align*}
  S^{-1} = I - P (D_{S}^{-1} + P^{T} P)^{-1} P^{T},
\end{align*}
reducing the problem of computing the inverse of $S$ to that of
computing the inverse of the reduced $\nprojs \times \nprojs$ matrix
$(D_{S}^{-1} + P^{T} P)$. This method for inverting $S$ was also used
in \cite{hasnip2006electronic} in the context of preconditioning in
ultrasoft computations.

\begin{figure}[h!]
  \centering
  \resizebox{.7\textwidth}{!}{\includegraphics{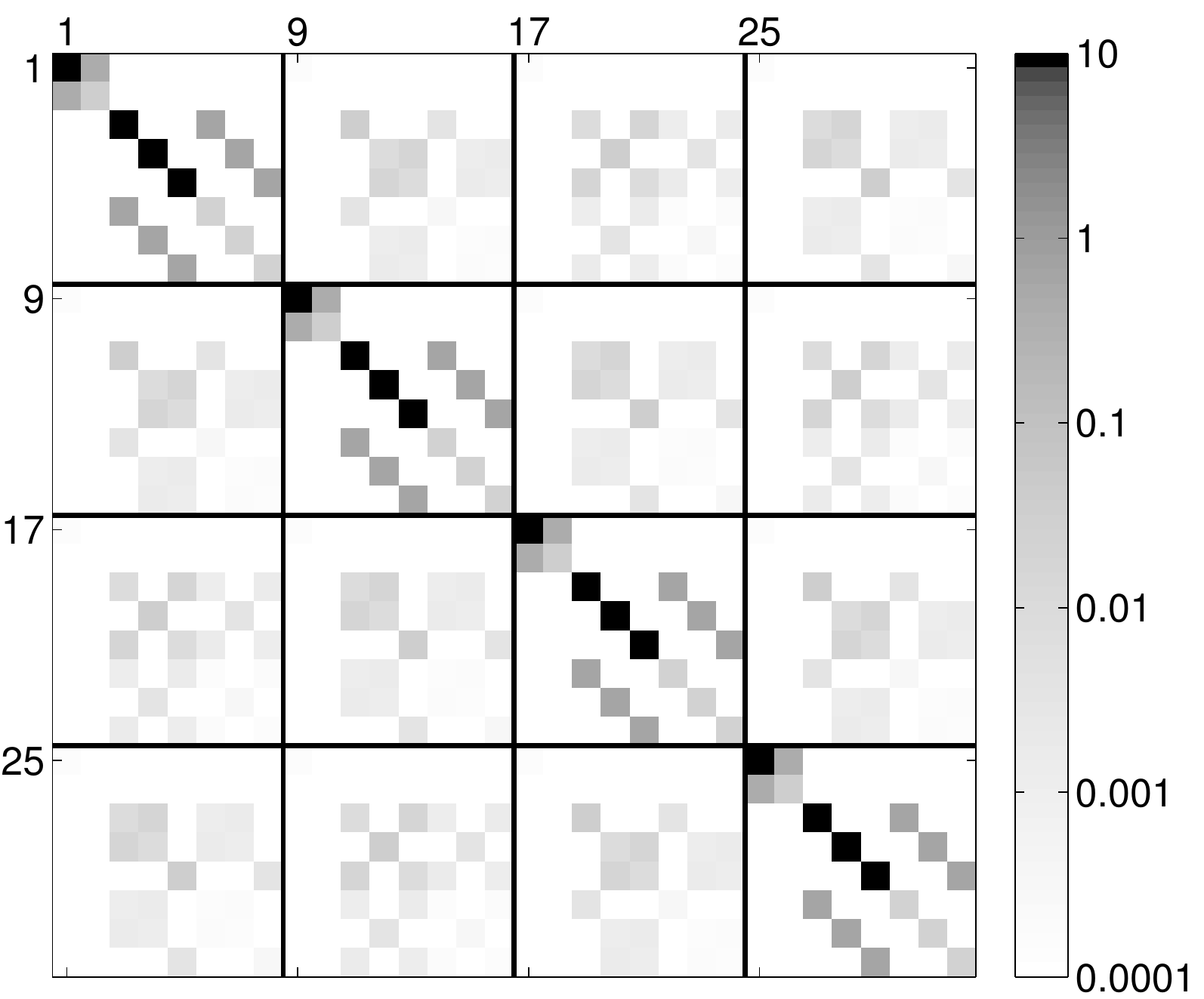}}
  \caption{Projectors overlap matrix $P^{T} P$ (logarithmic color
    scale) for a system of 4 aluminium atoms in a periodic
    box, with 8 projectors by atom. As an indication of the size of
    the overspill, denoting by $M$ the preconditioner obtained by
    keeping only the $8 \times 8$ diagonal blocks, the condition
    number of $M^{-1} (P^{T} P)$ was 2.64, and the preconditionned
    MINRES solver for the solution of $P^{T} P x = b$ with $b$ a
    random vector converged to machine precision in 7 iterations.}
  \label{fig:gram_projs}
\end{figure}

In PAW, the projectors are compactly supported in spheres centered
around the atoms. This leads us to expect that the matrix $(D_{S}^{-1}
+ P^{T} P)$ is block-diagonal, and therefore easy to invert. However,
the projectors $P$ are the discretization on the plane-wave basis of
the true PAW projectors. Because a function cannot be compactly
supported in both Fourier and real space, the plane-wave
discretization of the projectors will spill over the neighbouring PAW
spheres, and the Gram-matrix will have off-block diagonal entries (see
Figure \ref{fig:gram_projs}). This phenomenon is all the more
pronounced when the projectors are not smooth (and therefore have slow
Fourier-space decay), which is the case in many pseudopotentials
commonly used (often constructed by imposing matching conditions).

The result of this is that the matrix $(D_{S}^{-1} + P^{T} P)$ can not
be considered block-diagonal, or even sparse. While smaller than the
full matrix $S$, it is still too large to invert directly in large
systems. Therefore, we use an iterative solver, preconditionned by the
block-diagonal component of $(D_{S}^{-1} + P^{T} P)$. Since the
spillover phenomenon is relatively small, the preconditionner is a
very good approximation of the full matrix, and any iterative solver
converges to machine precision in a relatively modest number of
iterations. We used iterative refinement for its ease of
implementation, although any symmetric indefinite solver such as
MINRES could be used. In our tests, iterative refinement converged in
about 10 to 20 iterations, depending on the energy cutoff of plane
waves and the size of the PAW spheres. The cost of this inner
iterative solver is $O(\nprojs^{2})$, and therefore small compared to
the total cost $O(\nprojs \npw)$ of applying the overlap operator.



One inner iteration of CG or LOBPCG requires one multiplication by
$P^{T}$ and two by $P$ (one for $H$ and one for $S$). Naively
implemented, one inner iteration of the Chebyshev filter requires two
multiplications by $P^{T}$ and two by $P$ (one for $H$ and one for
$S^{-1}$). However, we can avoid the multiplication by $P^{T}$ for $H$
after the first iteration. Indeed, $P^{T} S^{-1} \psi$ can be written
as $P^{T} \psi - P^{T} P q$, where both $P^{T} \psi$ and $q$ have been
computed before. By precomputing the $\nprojs \times \nprojs$ Gram
matrix $P^{T} P$, this computation can be done in $O(\nprojs^{2})$
instead of the naive $O(\nprojs \npw)$.

Using this trick, the number of $O(\npw \times \nprojs)$ operations
for the application of a Chebyshev filter of degree $n$ is just one
more as the number of such operations that would be necessary for $n$
steps of a conjugate gradient algorithm. The iterative algorithm
described has an additional cost of $O(\nprojs^{2}) \ll O(\npw \times
\nprojs)$. In our tests, we found that this additional cost per
iteration compared to LOBPCG was largely compensated by the lack of
orthogonalization: therefore, one step of Chebyshev filtering is a
little faster than one step of LOBPCG.

\subsection{Parallelism}
We have implemented this algorithm in the ABINIT software using
MPI. The $\npw \times \nband$ eigenvector matrix is distributed on a
2D $\nppw \times \npband $ processors grid, in the same way as in
\cite{bottin2008large}. We apply the polynomial filter of degree
\nline, requiring communication inside the $\nppw$ processor group for
the FFT and the reductions needed for the nonlocal operator. Then, we
transpose the data to a $(\nppw \npband) \times 1$ grid (using the MPI
call \texttt{MPI\_ALLTOALL}), build the submatrices, distribute them
between the processors and perform a Rayleigh-Ritz procedure. Note
that, compared to LOBPCG, there is only one Rayleigh-Ritz per outer
(self-consistent) iteration.

All our computations are done on the Curie supercomputer installed at
the TGCC in France, a cluster of 16-core Intel processors with a total
of about 80,000 processors. We used the Intel MKL library for BLAS and
LAPACK dense linear algebra, and the ELPA library
\cite{auckenthaler2011parallel} for the dense eigenproblem in the
Rayleigh-Ritz step (in our tests, we found it was about twice as fast
as ScaLAPACK).

\subsection{Choice of the polynomial}
The choice of the polynomial degree \nline is a subtle matter,
requiring a balance between stability, speed and convergence.

First, a small degree results in many Rayleigh-Ritz steps, which is
detrimental to performance, and especially to parallelism. On the
other hand, if $\nline$ is too large, we will solve very accurately an
inaccurate problem, since the density $\rho$ is not yet
converged. Experience with the CG and LOBPCG algorithms have showed
that increasing $\nline$ above a moderate value (the default in ABINIT
is 4) does not speed up the self-consistent cycle. The same goes for
the Chebyshev filtering algorithm. More details are provided in
Section \ref{sec:numres}.

Finally, if the degree is too large, the Gram matrix $S_{\psi}$ will
be ill-conditionned, even if the columns of $\Psi$ are rescaled
beforehand. This leads to loss of precision (and, crucially, of
orthogonality). Therefore, we must have $T_{\nline}(\frac{\lambda_{1}
  - c}r) \ll 1/\varepsilon$, where $T_{n}$ is the Chebyshev polynomial
of degree $n$, and $\varepsilon \approx 10^{-16}$ is the machine
precision. In our tests, this was always the case except for large
values of $\nline$, of about $20$, and therefore this instability is
not an issue.

A key ingredient to the success of this algorithm is a good bracketing
of the unwanted part of the spectrum. The authors in \cite{seq_chebfi}
propose a few steps of the Lanczos algorithms to compute an upper
bound, but we simply use the upper bound $E_{\text{cut}}$ on the
kinetic energy of our system. It is not mathematically clear that this
is an upper bound of the operator $H = K + V$, since $V$ is not
non-positive, but we have found this to be true in all our numerical
tests.

To obtain an approximation of the smallest eigenvalue in the unwanted
part of the spectrum, we use the maximum Rayleigh quotient (always an
overestimation of the largest wanted eigenvalue
$\lambda_{\nband}$). We have found this to be more efficient than
using the largest Ritz value of the previous self-consistent iteration.

\subsection{Locking}
An important point for an effective implementation is the ability to
\textit{lock} converged eigenvectors, and not iterate on
them. Although it is far from clear what the optimal policy is in
terms of self-consistent convergence (how to optimize the number of iterations on
each band to obtain the lowest total running time), it is desirable to
stop the iterations prematurely in computations where a specified
accuracy on the wavefunctions is desired.

In the Chebyshev algorithm, this means adaptatively choosing the
degree of the polynomial, band per band. The problem is that there is
no simple way to obtain a measure of the error while applying the
polynomial: the vector being iterated on will become a combination of
all the eigenvectors, and the size of its residual is meaningless
before the Rayleigh-Ritz step. However, since an application of the
Chebyshev polynomial of degree $n$ enlarges the eigencomponent
associated with eigenvalue $\lambda_i$ by a factor
$T_{n}(\frac{\lambda_{i} - c}r)$, with the unwanted eigencomponents
multiplied by a factor of at most one, we can use the following
approximation (which becomes exact at convergence) for the residual
$r_i^n$ of band number $i$ with Rayleigh quotient $\lambda_i$ after
one full Chebyshev iteration (application of a Chebyshev polynomial of
degree n followed by a Rayleigh Ritz step)
\begin{align}
  \norm{r_{i}^n} \approx \frac{\norm{r_{i}}}{T_{n}(\frac{\lambda_{1} - c}r)},
\end{align}
where $r_i$ is the residual before the Chebyshev iteration (more
details can be found in \cite{saad1992numerical} and in references
therein). Using this estimate, we can choose a priori the polynomial
degree that will be needed to achieve a desired tolerance. This
prediction can also be useful for other purposes, such as providing
the user with an estimate of the progress of the computation.

Another issue is that using a polynomial of different degree for each
band leads to systematic load imbalance between the processors: since
the lower eigenvectors converge faster, the processors treating these
will have less work than those treating the slow-converging higher
eigenvectors. We avoid this by using a cyclic distribution of the
bands between the processors, so that each processor treats a mix of
low and high bands. This could be optimized further by redistributing
dynamically the bands so as to minimize the work imbalance, but we did
not implement this as the simple cyclic distribution led to a load
imbalance of less than $5\%$ in our tests.

By contrast, the LOBPCG algorithm suffers from incomplete locking when
a large number of processors is used, because the number of iteration
has to be the same for each band in a block. Because of the dependence
between the blocks, one cannot use a redistribution scheme such as in
the Chebyshev algorithm. To keep the comparaison fair between
Chebyshev and LOBPCG, we did not use any locking in the numerical
results presented here.
\section{Results}
\label{sec:numres}
\subsection{Non-self-consistent convergence}
As a first test, we study the non-self-consistent convergence of our solver,
meaning that we fix the Hamiltonian $H$ and focus on the linear
eigenvalue problem. Our test case is a system of $19$ atoms of Barium
titanate, with formula \ce{BaTiO3}, an insulator. We used an energy
cutoff of 20 hartrees, representative of standard computations. With
our PAW pseudopotential, there is a total of 77 totally filled
bands. We run three algorithms: CG, the classical conjugate gradient
of ref. \cite{payne1992iterative, kresse1996efficient}, the implementation
of LOBPCG in ABINIT \cite{bottin2008large}, and our Chebyshev
algorithm. The full-block version of LOBPCG ($\nblocks = 1$) was
used. In all cases, the parameter \nline, which controls the number of
inner iterations in all three algorithms, was set to 4. We monitor the
convergence of all eigenpairs using their residual $\norm{H \psi -
  \lambda S \psi}$.


\begin{figure}[h!]
  \centering
  {
    \begin{tikzpicture}
      \begin{axis}[xlabel = {Eigenvalue $\lambda$}, ylabel = {Number of iterations}, ymin=0,ymax=100, legend entries =
        {Chebyshev, CG, LOBPCG}, legend style = {at={(0.05,.95)}, anchor= north
          west}, legend cell align=left, \axisopt, xtick = {-2, -1.5, ..., 2}]
        \addplot[mark=\chebmarker,draw=chebcolor, thick] table [x index = 0, y index = 1 ] {data_100bands};
        \addplot[mark=diamond,draw=cgcolor, thick, mark size = 3pt] table [x index = 0, y index = 2 ] {data_100bands};
        \addplot[mark=\lobpcgmarker,draw=lobpcgcolor, thick, mark size = 3pt] table [x index = 0, y index = 3 ] {data_100bands};
      \end{axis}
    \end{tikzpicture}
  }
  \caption{Number of iterations to obtain a precision of $10^{-10}$,
    \ce{BaTiO3}, 100 bands.}
  \label{fig:cv_batio3_100bands}
\end{figure}
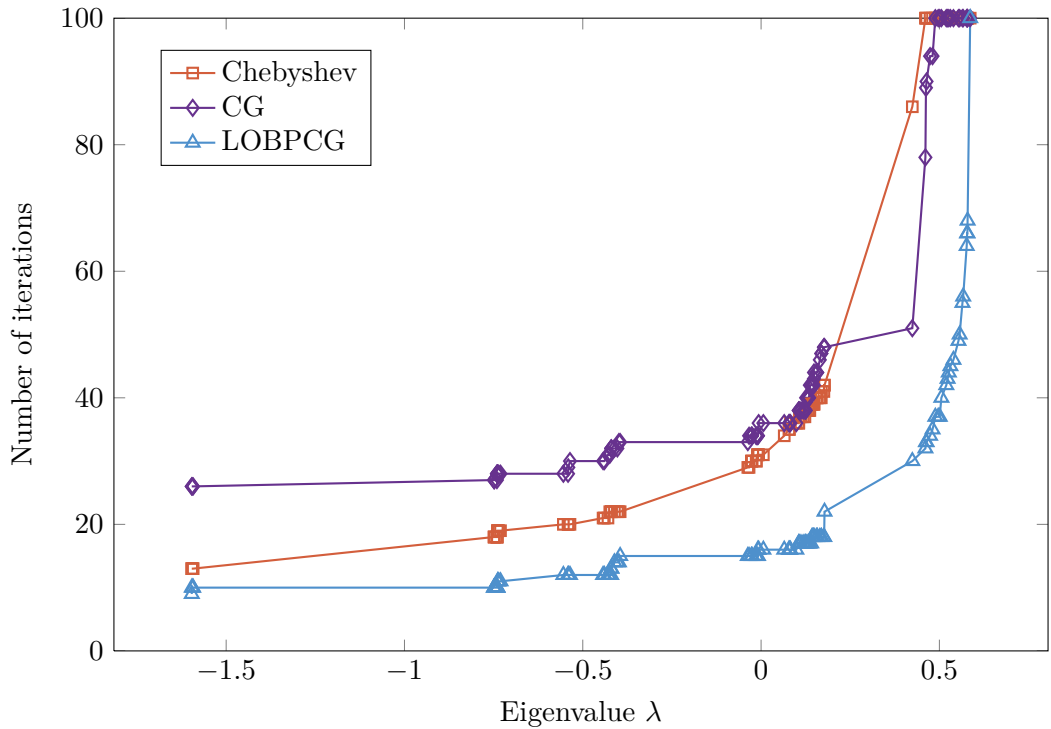

Figure \ref{fig:cv_batio3_100bands} displays the number of iterations
that was necessary for each eigenpair to attain an accuracy of
$10^{-10}$, using a total of 100 bands. We see that the Chebyshev
algorithm is very efficient towards the bottom of the spectrum,
outperforming the CG algorithm and even coming close to the full-block
LOBPCG algorithm. However, the situation degrades for the upper
eigenvalues, where the Chebyshev algorithm performs poorly, in part
due to the intrinsically poorer performance of Chebyshev algorithms
compared to Krylov methods, but in a large part due to the absence of
preconditionner. In tests not shown here, Chebyshev consistently
outperformed non-preconditionned CG and was competitive with
non-preconditionned LOBPCG except for the last eigenpairs.

\begin{figure}[h!]
  \centering
  {
    \begin{tikzpicture}
      \begin{axis}[xlabel = {Eigenvalue $\lambda$}, ylabel = {Number of iterations},ymin=0,ymax=100, legend entries =
        {Chebyshev, CG, LOBPCG}, legend style = {at={(0.05,.95)}, anchor= north
          west}, legend cell align=left, \axisopt, xtick = {-2, -1.5, ..., 2}]
        \addplot[mark=\chebmarker,draw=chebcolor, thick] table [x index = 0, y index = 1 ] {data_200bands};
        \addplot[mark=diamond,draw=cgcolor, mark size = 3, thick] table [x index = 0, y index = 2 ] {data_200bands};
        \addplot[mark=\lobpcgmarker,draw=lobpcgcolor, mark size = 3, thick] table [x index = 0, y index = 3 ] {data_200bands};
      \end{axis}
    \end{tikzpicture}
  }
  \caption{Number of iterations to get to obtain a precision of $10^{-10}$,
    \ce{BaTiO3}, 200 bands.}
  \label{fig:cv_batio3_200bands}
\end{figure}
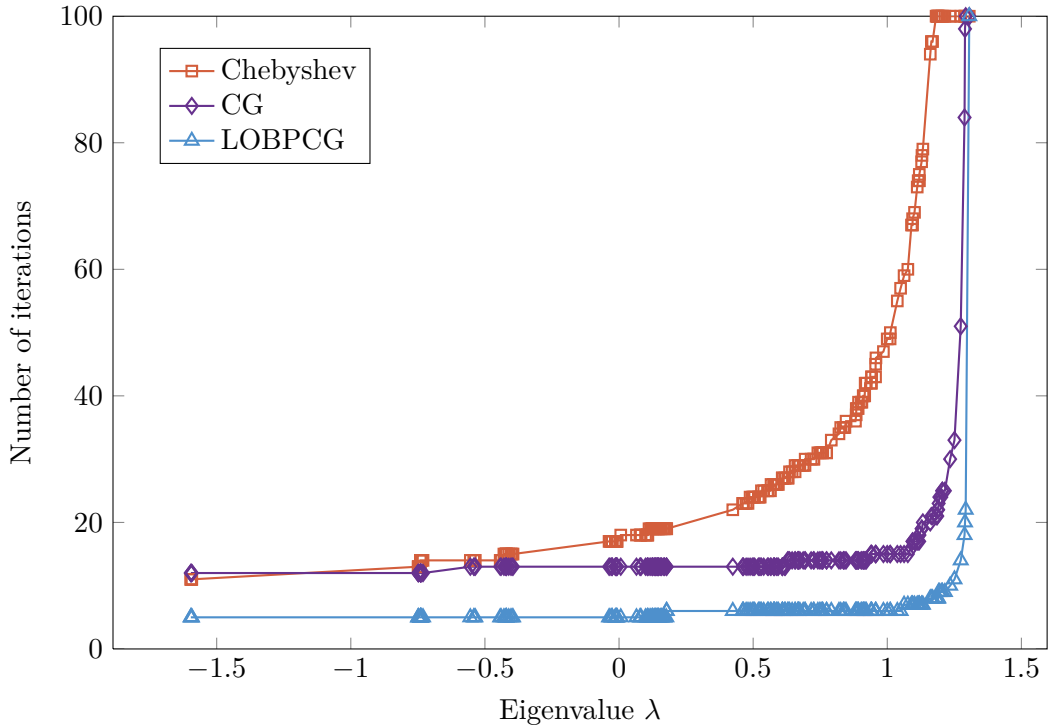

Figure \ref{fig:cv_batio3_200bands} shows the exact same computation,
but with 200 bands. First, note that increasing the number of bands
yields improved convergence rates: the eigenpairs near $0.5$ now converge in
about 30 iterations for Chebyshev, whereas they did not converge in
100 iterations before. The inclusion of a large number of bands in the
computation (200, compared with a total dimension of $\npw \approx
7000$) also greatly enhances the effectiveness of the CG algorithm,
although we do not fully understand this effect. In this situation,
the Chebyshev algorithm is not competitive.

We also note that the performance of the Chebyshev algorithm degrades
like $1/\sqrt{E_{\text{cut}}}$ as the energy cutoff is increased,
whereas LOBPCG and CG, thanks to their preconditionning, only show a
moderate increase in the number of iterations.



\subsection{Self-consistent convergence}
We now study the impact of the linear solver on the self-consistent cycle, and on
the overall efficiency. Our tests are on a crystal of 256 atoms of
Titanium. The partial occupation scheme used leads to about 1300 fully
occupied bands and 500 partially occupied ones. We performed our tests
with a total of 2048 bands.

\begin{figure}[h!]
  \centering
  \begin{tikzpicture}
    \begin{semilogyaxis}[xlabel = {$n$}, ylabel = {Residual on $V$}, ymin=1e-10, ymax = 1e4, legend entries =
      {Chebyshev \nline 4, Chebyshev \nline 8, LOBPCG \nline 4, LOBPCG \nline 8}, legend style = {at={(0.95,.95)}, anchor= north
        east}, legend cell align=left, \axisopt]
      \addplot[mark=\chebmarker,draw=chebcolor, mark size = 3, thick] table [x index = 0, y index = 4 ] {data_cheb_nline4};
      \addplot[mark=\chebmarker*,draw=chebcolor, mark size = 3, mark
      options = {fill=chebcolor!60}, thick] table [x index = 0, y index = 4 ] {data_cheb_nline8};
      \addplot[mark=\lobpcgmarker,draw=lobpcgcolor, mark size = 3, thick ] table [x index = 0, y index = 4 ]  {data_lobpcg_nline4};
      \addplot[mark=\lobpcgmarker*,draw=lobpcgcolor, mark size = 3, thick,
      mark options = {fill=lobpcgcolor!60}] table [x index = 0, y index = 4 ] {data_lobpcg_nline8};
    \end{semilogyaxis}
  \end{tikzpicture}

  \caption{Self-consistent convergence. The blocksize for LOBPCG was 128.}
  \label{fig:cv_ti256}
\end{figure}
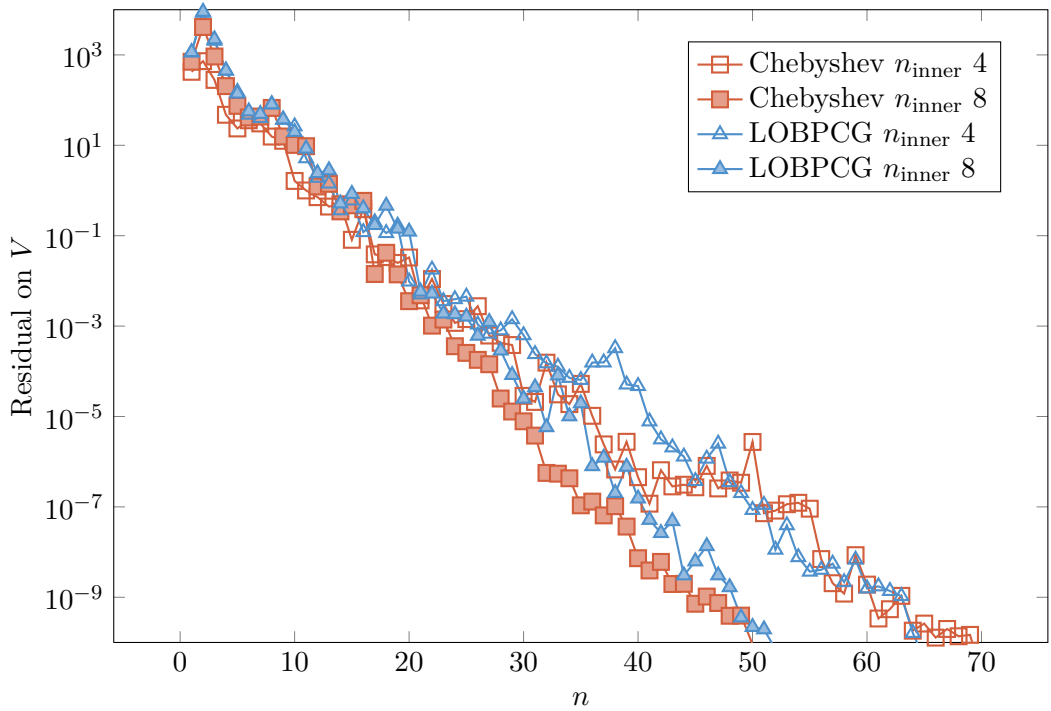

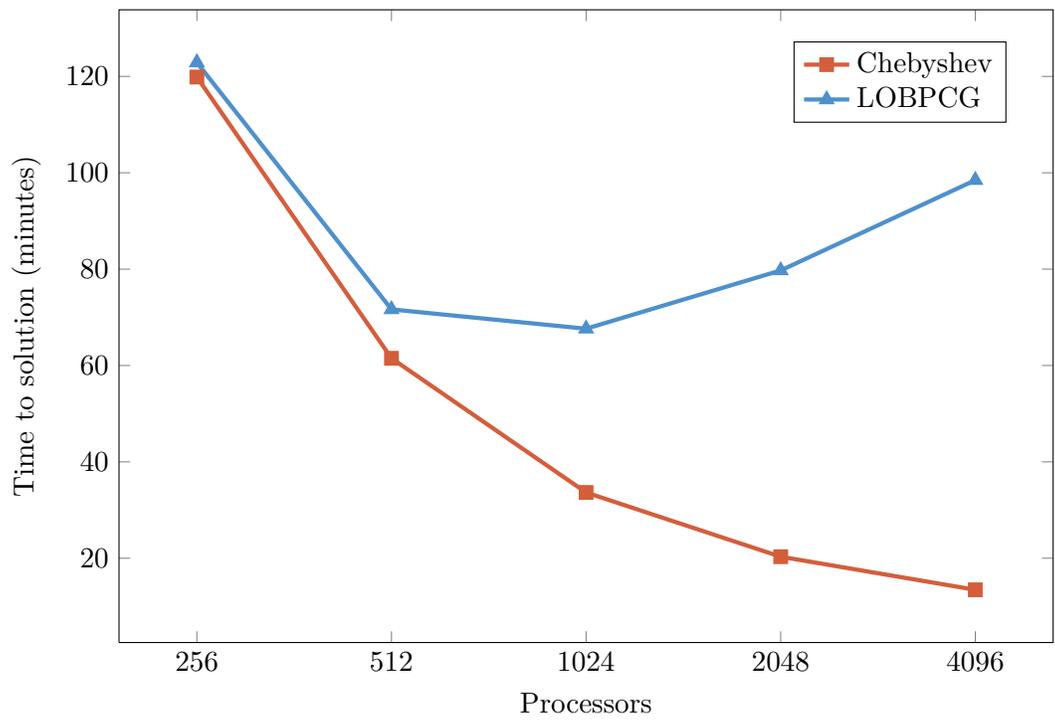
\begin{figure}[h!]
  \centering
  \begin{tikzpicture}
    \begin{semilogxaxis}[xlabel = {Processors}, ylabel = {Time to solution (minutes)}, legend entries =
      {Chebyshev, LOBPCG}, legend style = {at={(0.95,.95)}, anchor= north
        east}, legend cell align=left, \axisopt,
      xtick=data, log basis x = 2, log basis y = 2,xticklabels = {256, 512, 1024,
        2048, 4096, 8192},
      ]
      \addplot[mark=\chebmarker*,draw=chebcolor, ultra thick, mark
      options = {fill=chebcolor}] table [x index = 0, y expr = \thisrowno{1}/60 ] {data_tts};
      \addplot[mark=\lobpcgmarker*,draw=lobpcgcolor, ultra thick, mark options = {fill=lobpcgcolor}] table [x index = 0, y expr = \thisrowno{2}/60 ] {data_tts};
    \end{semilogxaxis}
  \end{tikzpicture}

  \caption{Total time to solution. \nline was fixed to 4, and \nppw to 32. The total number of
    iterations was 70 for Chebyshev. It varied from 65 to 55 for LOBPCG,
    as the blocksize was increased from 32 on 256 processors to 512
    on 4096.}
  \label{fig:time_to_solution}
\end{figure}

The computations were stopped when the residual on the potential went
below $10^{-10}$. We report the convergence for \nline equal
to 4 and 8 in Figure \ref{fig:cv_ti256}. The results show that
Chebyshev and LOBPCG are competitive on this system. The superior
parallel performance of the Chebyshev algorithm yields large speedups
when using more processors, as can be seen in Figure
\ref{fig:time_to_solution}. In this case, taking the best time among
all processor numbers yields a total time of about 15 minutes on 4096
processors for Chebyshev compared to more than an hour with 1024
processors for LOBPCG.

\clearpage

\subsection{Scalability}
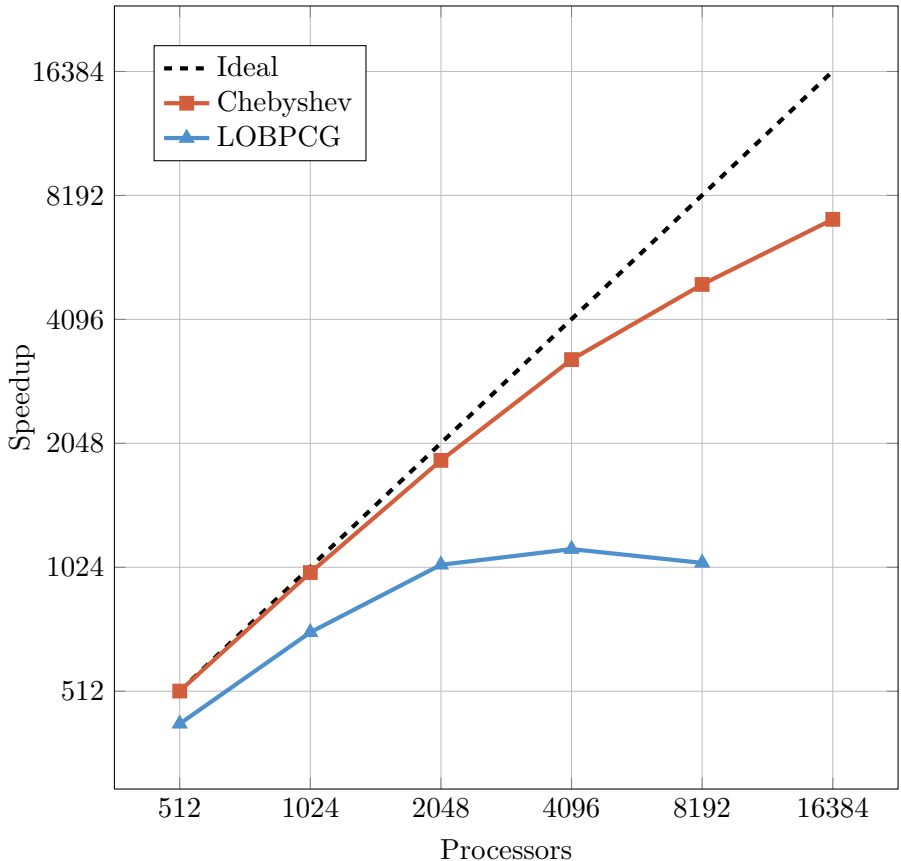
\begin{figure}[h!]
  \centering
  \begin{tikzpicture}
    \begin{loglogaxis}[xlabel = {Processors}, ylabel = {Speedup}, legend entries =
      {Ideal, Chebyshev, LOBPCG}, legend style = {at={(0.05,.95)}, anchor= north
        west}, legend cell align=left, height=12cm, width=12cm,
      xtick=data, log basis x = 2, log basis y = 2, xticklabels = {512, 1024,
        2048, 4096, 8192, 16384}, yticklabels = {512, 512, 1024, 
        2048, 4096, 8192, 16384},
      grid = major
      ]
      \addplot[draw=black,mark=none, dashed, ultra thick] table [x index = 0, y index = 0 ] {data_timings};
      \addplot[mark=\chebmarker*,draw=chebcolor, ultra thick, mark options = {fill=chebcolor}] table [x index = 0, y index = 1 ] {data_timings};
      \addplot[mark=\lobpcgmarker*,draw=lobpcgcolor, ultra thick, mark options = {fill=lobpcgcolor}] table [x index = 0, y index = 2 ] {data_timings};
    \end{loglogaxis}
  \end{tikzpicture}

  \caption{Speedups for the Chebyshev and LOBPCG method, \ce{Ti_{512}}.}
  \label{fig:speedup}
\end{figure}

We now study more precisely the parallel scalability of our algorithms
on a cristal of 512 atoms of Titanium, with a total of 4096 bands. We
chose $\nppw = 64$. As before, we chose \nline $ = 4$. We
measured the average running time of a single iteration. We began our
measurements at 512 processors, and timed the individual routines. The
speedups of Figure \ref{fig:speedup} are obtained with reference to a
base case extrapolated by substracting the time spent in
communications from the total time.

\begin{figure}[h!]
  \centering
  \begin{tikzpicture}
    \begin{loglogaxis}[xlabel = {Processors}, ylabel = {Total time (s)}, legend entries =
      {Total, Hamiltonian, Rayleigh-Ritz, Others}, legend style = {at={(0.95,.95)}, anchor= north
        east}, legend cell align=left, \axisopt,
      xtick=data, log basis x = 2,xticklabels = {512, 1024,
        2048, 4096, 8192, 16384},
      ]
      \addplot+[draw=black, ultra thick, mark options = {fill=black}] table [x index = 0, y index = 4 ] {data_breakdown};
      \addplot+[draw=linea, ultra thick, mark options = {fill=linea}] table [x index = 0, y index = 1 ] {data_breakdown};
      \addplot+[draw=lineb, mark=diamond*, ultra thick, mark options = {fill=lineb}] table [x index = 0, y index = 2 ] {data_breakdown};
      \addplot+[draw=line06, mark = triangle*, ultra thick, mark options = {fill=line06}] table [x index = 0, y index = 3 ] {data_breakdown};
    \end{loglogaxis}
  \end{tikzpicture}

  \caption{Breakdown of one step of the Chebyshev method, \ce{Ti_{512}}.}
  \label{fig:breakdown}
\end{figure}
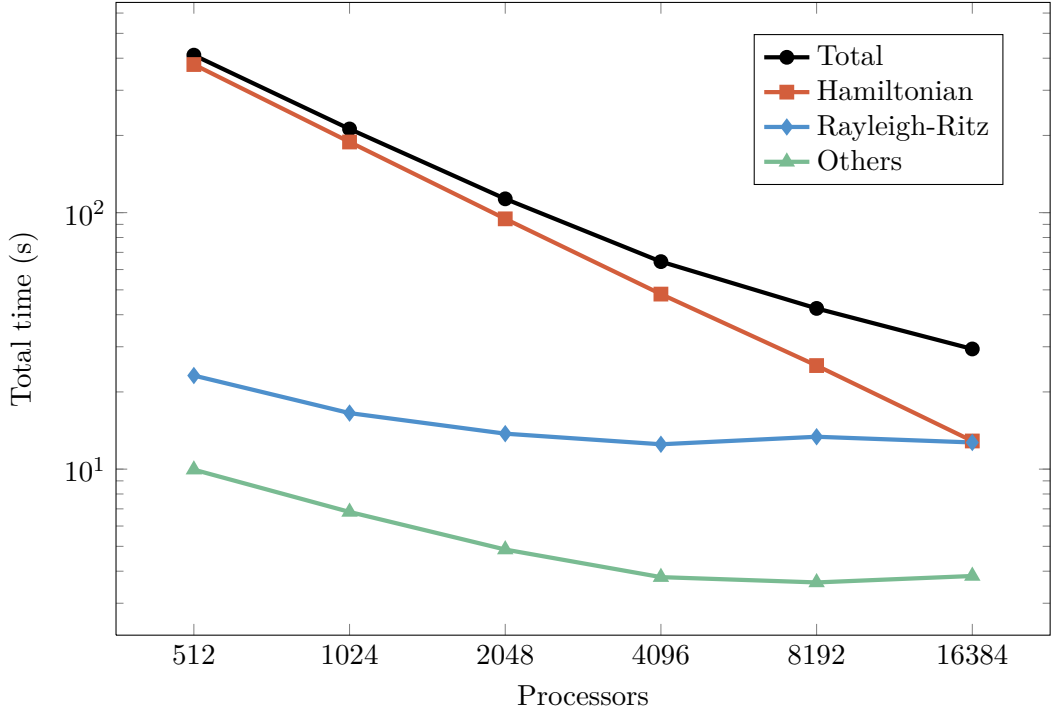

The scalability for the Chebyshev algorithm is again much better,
still scaling at 16384 processors for the \ce{Ti_{512}} crystal when LOBPCG
saturates at 2048. Figure \ref{fig:breakdown} shows the breakdown of
a step of the Chebyshev method. While the Hamiltonian application
scales perfectly, as expected, the Rayleigh-Ritz step saturates very
quickly and goes from negligible at 512 to being as costly as the
Hamiltonian application at 16384 processors.

\section{Conclusion}
Using a Woodbury decomposition of the PAW overlap matrix, we extended
the Chebyshev filtering algorithm of \cite{seq_chebfi,
  parallel_chebfi} to a generalized eigenproblem, and implemented it
in the ABINIT software. Comparisons with the current implementation,
based on the LOBPCG algorithm, show that its convergence properties
are competitive for some systems, although it proves slower for
others, due to its lack of preconditioning. Because it needs much less
Rayleigh-Ritz steps, it is able to achieve much greater parallel
speedups, and scale into the tens of thousands of processors.

This scaling behavior is acceptable for current generations of
machines, where it is rare to be able to use more than 10,000
cores. However, exascale computations will only be possible with the
help of algorithms that avoid global Rayleigh-Ritz steps. For
plane-wave DFT, the only competitive algorithm seems to be the
spectrum slicing algorithm of \cite{slicing}, but the high-degree
polynomials it uses render it uncompetitive for all but extremely
large systems. More research is needed to be able to develop
alternatives.

\section*{Acknowledgements}
We wish to thank Laurent Colombet for helpful discussions on HPC
issues. Access to the Curie supercomputer was provided through the
Centre de Calcul Recherche et Technologie (CCRT).
\bibliographystyle{abbrv}
\bibliography{refs}

\end{document}